# Two distinct methods to evaluate graphene relaxation time and mobility in Boltzmann diffusive transport, considering ionized impurity scattering and Thomas-Fermi screening

Yi Zhan[*]

*University of Southern California, Los Angeles, CA 90089, USA*

**ABSTRACT**

Boltzmann diffusive transport including relaxation time and mobility in graphene limited by ionized impurity scattering is investigated. The relaxation time is evaluated with two different methods, first one directly use Boltzmann transport equation via scattering matrix, second one is based on scattering cross-section. Two methods yield the same relaxation time results for graphene. Assume linear Thomas-Fermi screening and a reasonable electron carrier density, relaxation time $\tau_m = 17 ps$ and mobility $\mu \approx 1.43 \times 10^6 cm^2 V^{-1} s^{-1}$ can be calculated and plotted, which demonstrates graphene is a very promising high mobility material for ultra fast electronic device applications.

*Author to whom correspondence should be addressed: yizhan@usc.edu

## I. Introduction

Graphene as one of monolayer two-dimensional (2D) materials has attracted a lot of research interests since it was first isolated in 2004 [1][2]. One of the most superior electronic property in graphene is the high carrier mobility, which make graphene a promising candidate for next generation high speed electronic devices and energy devices [3-5] The unique massless Dirac fermion structure in graphene also makes graphene transport special, compared to conventional 2D electron gas. In this paper, two distinct methods are used to derive and evaluate relaxation time in graphene diffusive transport. One is directly from Boltzmann transport equation; the other way is based on scattering cross-section. The results are shown to be the same. Linear Thomas Fermi screening, zero temperature and ionized impurity scattering are used to simplify the calculation process and demonstrate the physics. In the end, potential energy, relaxation time and carrier mobility are calculated, plotted and discussed.

In additional to unique ballistic transport [6-8], diffusive transport occurs in graphene when medium dimension (e.g. device length) is much longer than carrier mean free path, which is a more important transport mechanism for electronic device applications. The Boltzmann transport equation (BTE) is used as a standard way to study the diffusive transport [9-13].

$$\frac{\partial f}{\partial t} + \vec{v} \cdot \nabla_{\vec{r}} f - \frac{e\vec{\xi}}{\hbar} \cdot \nabla_{\vec{k}} f = \left(\frac{\partial f}{\partial t}\right)_{scatter} \quad (1)$$

$\vec{v}$ is the carrier velocity, $f$ is the number of particles, $\vec{\xi}$ is the applied electric field. In steady state without $\vec{r}$ dependence, Eq. 1 can be simplified into a reduced Boltzmann equation. Applying the relaxation time approximation and considering a low field $|\vec{\xi}|$ for linear response, we arrive at[9]

$$-\frac{e\vec{\xi}}{\hbar} \cdot \nabla_{\vec{k}} f = \left(\frac{\partial f}{\partial t}\right)_{scatter} = -\frac{f(\vec{k}) - f_0(\vec{k})}{\tau(\vec{k})} \quad (2)$$

$$f(\vec{k}) = f_0(\vec{k}) + \frac{e}{\hbar}[\vec{\xi} \cdot \nabla_{\vec{k}} f_0(\vec{k})]\tau(\vec{k}) \quad (3)$$

The transition rate from state $|\vec{k}\rangle$ to state $|\vec{k}'\rangle$ is denoted as $P_{\vec{k},\vec{k}'}$, and can be written down by Fermi's golden rule

$$P_{\vec{k},\vec{k}'} = \frac{2\pi}{\hbar}|\langle\vec{k}'|H|\vec{k}\rangle|^2 n_{imp}\delta[E(\vec{k}') - E(\vec{k})] \quad (4)$$

$n_{imp}$ is the impurity concentration, $H$ is the interactive Hamiltonian. For elastic scattering mechanisms such as ionized impurity scattering, energy is conserved and $P_{\vec{k},\vec{k}'} = P_{\vec{k}',\vec{k}}$ can be satisfied. Applying the famous detailed balanced equation, [i] we arrive with Eq. 5

$$\left(\frac{\partial f}{\partial t}\right)_{scatter} = \frac{1}{A}\sum_{\vec{k}'} P_{\vec{k},\vec{k}'}\{[1-f(\vec{k})]P_{\vec{k},\vec{k}'}f(\vec{k}') - [1-f(\vec{k}')]P_{\vec{k}',\vec{k}}f(\vec{k})\}$$

$$= \sum_{\vec{k}'} P_{\vec{k},\vec{k}'}[f(\vec{k}') - f(\vec{k})] \quad (5)$$

Substitute Eq. 5 into Eq. 2, after some mathematical rearrangements, the momentum relaxation time for isotropic material like graphene can be written down as

$$\frac{1}{\tau_m(\vec{k})} = \frac{1}{(2\pi)^2}\int_{all\ \vec{k}'} d^2k'\ P_{\vec{k},\vec{k}'}(1-cos\theta) \quad (6)$$

scattering angle $\theta = \theta_{\vec{k}'} - \theta_{\vec{k}}$, and $\theta_{\vec{k}}$ is the angle for incoming state wave vector $|\vec{k}\rangle$. Since graphene is an isotropic material, $\tau_m = \tau_m(|\vec{k}|)$. In other words, $\tau_m$ is independent of the direction of $|\vec{k}\rangle$ for the isotropic system. While for anisotropic materials such as 2D black phosphorus, the relaxation time is much more complicated shown as Eq.7, more information on Boltzmann transport and and relaxation time derivation can be found in reference [14-15].

$$\frac{1}{\tau_m(\hat{\xi},\vec{k}_i)} = \frac{1}{(2\pi)^2}\int_{all\ \vec{k}_j} d^2k_j\ P_{\vec{k}_i,\vec{k}_j}\left\{1 - \frac{[\hat{\xi}\cdot\vec{v}(\vec{k}_j)]\tau_m(\hat{\xi},\vec{k}_j)}{[\hat{\xi}\cdot\vec{v}(\vec{k}_i)]\tau_m(\hat{\xi},\vec{k}_i)}\right\} \quad (7)$$

We start with schematic model in Fig. 1 to study graphene diffusive transport properties. To be specific, we consider an atomically thin, infinite sheet of graphene on a $SiO_2$ insulator with a non-magnetic, ionized impurity that is located a distance $z_0$ below the graphene- $SiO_2$ interface. The modeled geometry is cylindrically symmetric. We focus on electrons in this section and thus graphene is *n*-type doped.

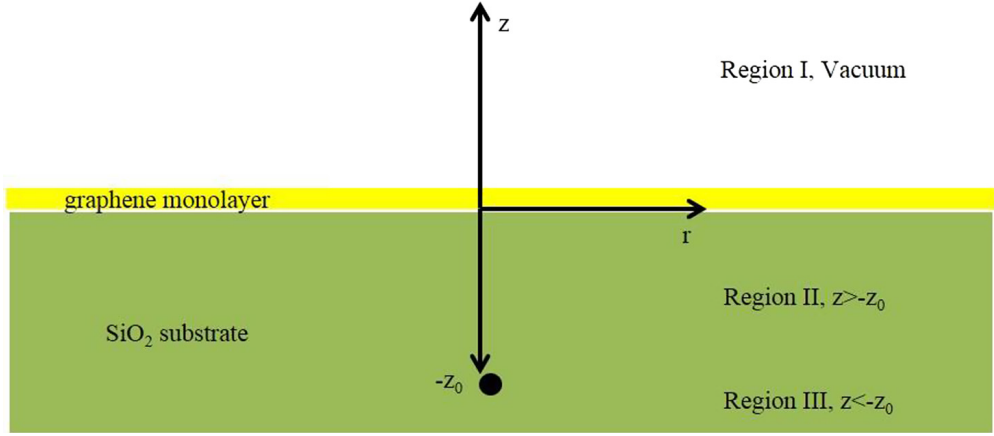

Fig. 1. Schematic modeling of a back-gated graphene device (e.g. Fig. 2.5) with out-of-plane impurity. The graphene is located at the $z = 0$ plane and the impurity charge is located at $-z_0$ ($z_0 > 0$).

Among all the scattering mechanisms, we mainly consider ionized impurity scattering (in some literature also called charged impurity interaction, or Coulomb impurity interaction) because (1) ionized impurity scattering is the dominating mechanism in low temperature transport (e.g. $T < 100K$); (2) ionized impurity scattering is relatively straightforward. Once we get familiar with the calculation procedures, we can extend our theory to other scattering mechanisms by switching the interaction Hamiltonian $H$ in Eq. 4.

## II. Relation time evaluation using scattering matrix

For ionized impurity interaction, the scattering potential energy $U$ at $(\vec{r}, z)$ can be written as

$$U(\vec{r}, z) = \frac{e^2}{\kappa_1}\left(\frac{1}{R_1} - \frac{\kappa_2 - \kappa_1}{\kappa_2 + \kappa_1}\frac{1}{R_2}\right) \quad (8)$$

$R_1 = \sqrt{r^2 + (z_0 + z)^2}$ and $R_2 = \sqrt{r^2 + (z_0 - z)^2}$ is due to geometry in Fig. 1, the factor $\frac{\kappa_2 - \kappa_1}{\kappa_2 + \kappa_1}$ is due to image charge effect. For the graphene monolayer at $z = 0$, plugging $z = 0$ into $R_1$ and $R_2$, Eq. 8 can be simplified as

$$U(\vec{r}) = U(r) = \frac{e^2}{\kappa\sqrt{r^2 + z_0^2}} \quad (9)$$

The effective $\kappa = \frac{\kappa_1 + \kappa_2}{2} = \frac{1}{2}(3.9 + 1) \approx 2.5$, with $\kappa_1$ and $\kappa_2$ are dielectric constants of SiO$_2$ and vacuum, respectively. For ionized impurity scattering limited graphene, Hamiltionian $H$ is the Coulomb potential energy $U(r)$.

The two normalized "pseudo-spin" [11, 18] eigenstates in graphene can be written as

$$|\vec{k}\rangle = \frac{1}{\sqrt{2}}\begin{pmatrix} e^{-i\frac{\theta_{\vec{k}}}{2}} \\ e^{i\frac{\theta_{\vec{k}}}{2}} \end{pmatrix} e^{i\vec{k}\cdot\vec{r}} \quad (10)$$

The scattering matrix $\langle \vec{k}'|H|\vec{k}\rangle$ be written as

$$\langle \vec{k}'|H|\vec{k}\rangle = \frac{1}{2}\begin{pmatrix} e^{i\frac{\theta_{\vec{k}'}}{2}} & e^{-i\frac{\theta_{\vec{k}'}}{2}} \end{pmatrix}\begin{pmatrix} e^{-i\frac{\theta_{\vec{k}}}{2}} \\ e^{i\frac{\theta_{\vec{k}}}{2}} \end{pmatrix} \int d^2 r\, e^{i(\vec{k}'-\vec{k})\cdot\vec{r}} \frac{e^2}{\kappa\sqrt{r^2 + z_0^2}} \quad (11)$$

We define the transfer wave vector $\vec{q} = \vec{k}' - \vec{k}$, and the scattering matrix becomes

$$\langle \vec{k}'|H|\vec{k}\rangle = \frac{e^2}{\kappa}\cos\frac{\theta}{2}\int_0^\infty dr\, \frac{r}{\sqrt{r^2 + z_0^2}}\int_0^{2\pi} d\theta\, e^{iqr\cos\theta} \quad (12)$$

The integrals in Eq. 12 can be done analytically with following two integral identities. It is worth mentioning that Eq. 13a and 13b are very commonly used in 2D materials transport theory.

$$\int_0^{2\pi} d\theta\, e^{iqr\cos\theta} = 2\pi J_0(qr) \quad (13a)$$

$$\int_0^\infty dr\, \frac{r}{\sqrt{r^2 + z_0^2}} J_0(qr) = \frac{e^{-qz_0}}{q} \quad (13b)$$

$J_0$ is the Bessel function of order zero. The potential energy in reciprocal space, $\hat{U}(\vec{q})$, can be solved using Fourier transform as shown in Eq. 14a. (a hat notation is used to distinguish real space and Fourier space). Subsequently, the scattering matrix $\langle \vec{k}'|H|\vec{k}\rangle$ can be written as Eq. 14b.

$$\hat{U}(\vec{q}) = \hat{U}(q) = \int d^2 r\, e^{i\vec{q}\cdot\vec{r}} U(r) = \frac{2\pi e^2}{\kappa}\frac{e^{-qz_0}}{q} \quad (14a)$$

$$\langle \vec{k}'|H|\vec{k}\rangle = \hat{U}(q)\cos\frac{\theta}{2} \quad (14b)$$

It is worth noticing that calculating the scattering matrix in Eq. 11-12 includes the same approach as doing the Fourier transform. Moreover, we find that both $U(r)$ and $\hat{U}(q)$ only depends on the magnitude of $r = |\vec{r}|$ or $q = |\vec{q}|$, which is true for graphene since it is an isotropic material.

Before we proceed to evaluate the diffusive transport in graphene due to ionized impurity scattering, we also need to consider the effect of electron screening. We start with a simple linear Thomas-Fermi screening. Considering the screening effect, the $\hat{U}(q)$ in Eq. 14 is modified to be

$$\widehat{U}(q) = \frac{2\pi e^2}{\kappa} \frac{e^{-qz_0}}{q+q_s} = \frac{2\pi e^2 e^{-qz_0}}{q\kappa + 2\pi e^2 aE} \tag{15}$$

Within the linear Thomas-Fermi screening, $q_s$ is given by $q_s = \frac{2\pi e^2}{\kappa} g(E_F)$. $g(E) = \frac{2}{\pi(\hbar v_F)^2} E$, $a = \frac{2}{\pi(\hbar v_F)^2}$, graphene carrier density $n = \frac{1}{2} aE_F^2$ at zero temperature. Finally, we arrive with the following scattering matrix

$$\langle \vec{k}'|H|\vec{k}\rangle = \frac{2\pi e^2}{\kappa} \frac{e^{-qz_0}}{q + \frac{2\pi e^2}{\kappa} g(E_F)} \cos\frac{\theta}{2} = \frac{2\pi e^2 e^{-qz_0}}{q\kappa + 2\pi e^2 aE_F} \cos\frac{\theta}{2} \tag{16}$$

Substituting Eq. 16 into the transition rate Eq. 4, $\cos^2\frac{\theta}{2} = \frac{1+\cos\theta}{2}$

$$P_{\vec{k},\vec{k}'} = \frac{2\pi}{\hbar} |\widehat{U}(q)|^2 \left(\frac{1+\cos\theta}{2}\right) n_{imp} \delta[E(\vec{k}') - E(\vec{k})] \tag{17}$$

Using Eq. 6, the momentum relaxation time $\tau_m$ can be written as

$$\frac{1}{\tau_m} = \frac{1}{(2\pi)^2} \frac{\pi n_{imp}}{\hbar} \int_{all\ \vec{k}'} d^2k' |\widehat{U}(q)|^2 (1+\cos\theta)(1-\cos\theta) \delta[E(\vec{k}') - E(\vec{k})] \tag{18}$$

Since energy disperson $E(\vec{k}) = \hbar v_F k$ for graphene, carrier Fermi velocity $v_F = 10^8 cm/s$ $\delta[E(\vec{k}') - E(\vec{k})] = \frac{\delta(k'-k)}{\hbar v_F}$, and $q = 2k\sin\frac{\theta}{2} = \frac{2E}{\hbar v_F}\sin\frac{\theta}{2}$ can be found in Fig. 2 for isotropic elastic scattering process, the momentum relaxation time calculated by Eq. 18 can be simplified to

$$\frac{1}{\tau_m} = \frac{n_{imp}}{4\pi\hbar} \int dk' k' \frac{\delta(k'-k)}{\hbar v_F} \int_0^{2\pi} d\theta |\widehat{U}(q)\sin\theta|^2$$

$$= \frac{n_{imp} E}{4\pi\hbar^3 v_F^2} \int_0^{2\pi} d\theta |\widehat{U}(q)\sin\theta|^2 \tag{19}$$

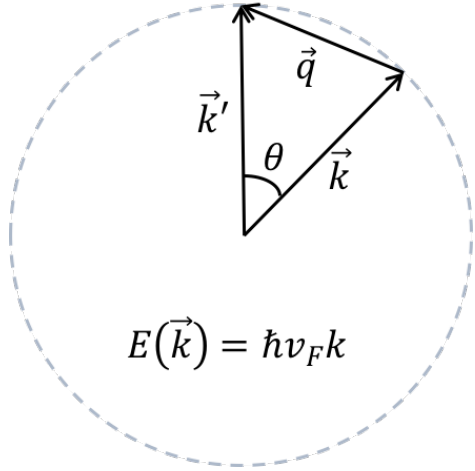

Fig. 2. Schematic for transfer vector $\vec{q} = \vec{k}' - \vec{k}$.

Now we have achieved relaxation time expression. The second term $\int_0^{2\pi} d\theta |\widehat{U}(q)sin\theta|^2$ in Eq. 19 is an integral which is solely determined by $E$, and can be calculated numerically. We assert that for isotropic material such as graphene, $\tau_m$ is indeed determined by carrier energy only, i.e. independent of incoming state direction.

### III. Relation time evaluation using scattering cross-section

Another way to derivate the relaxation time equation in Eq. 19 and evaluate the ionized impurity scattering in graphene via the differential scattering cross-section $\frac{d\sigma}{d\theta}\big|_{k\rightarrow k'}$ and scattering amplitude $f_B(\theta)$. In the lowest order Born approximation, the scattering amplitude for graphene can be found in Novikov [19].

$$f_B(\theta) = -\frac{1}{\hbar v_F}\sqrt{\frac{k}{8\pi}}\widehat{U}(q)(1 + e^{-i\theta}) \tag{20}$$

Considering $(1 + e^{-i\theta})^2 = 4cos^2\frac{\theta}{2}e^{-i\theta}$, differential scattering cross-section become [20]

$$\frac{d\sigma}{d\theta}\bigg|_{k\rightarrow k'} = |f_B(\theta)|^2 = \frac{E}{2\pi(\hbar v_F)^3}|\widehat{U}(q)|^2 cos^2\frac{\theta}{2} \tag{21}$$

Alternatively, $\frac{d\sigma}{d\theta}\big|_{k\rightarrow k'}$ can also be related with the scattering matrix $\langle\vec{k}'|H|\vec{k}\rangle$

$$\frac{d\sigma}{d\theta}\bigg|_{k\rightarrow k'} = \frac{g_{E(\vec{k}')}}{4\hbar v_F}|\langle\vec{k}'|H|\vec{k}\rangle|^2 \tag{22}$$

$g_{E(\vec{k}')}$ is the DOS for final state $|\vec{k}'\rangle$, i.e. $g_{E(\vec{k}')} = \frac{2}{\pi(\hbar v_F)^2}E$. Plugging the scattering matrix (Eq. 14b) in Eq. 22 and we obtain same expression as Eq. 21

$$\left.\frac{d\sigma}{d\theta}\right|_{k\to k'} = \frac{E}{2\pi(\hbar v_F)^3}|\hat{U}(q)|^2 \cos^2\frac{\theta}{2} \qquad (23)$$

The momentum relaxation time can be written as [10],

$$\frac{1}{\tau_m} = n_{imp} v_F \sigma(E) \qquad (24)$$

$\sigma(E)$ is the total scattering cross-section, which can be calculated by integrating $\left.\frac{d\sigma}{d\theta}\right|_{k\to k'}$,

$$\sigma(E) = \int d\theta \left.\frac{d\sigma}{d\theta}\right|_{k\to k'} (1-\cos\theta)$$

$$= \frac{E}{2\pi(\hbar v_F)^3} \int d\theta |\hat{U}(q)|^2 \cos^2\frac{\theta}{2}(1-\cos\theta) \qquad (25)$$

The $(1-\cos\theta)$ term in Eq. 25 corresponds to the isotropic momentum (or potential) relaxation process in graphene. This term appears in Eq. 6, too. Subsequently the relaxation time $\tau_m$ can be recovered,

$$\frac{1}{\tau_m} = n_{imp} v_F \sigma(E) = \frac{n_{imp} E}{4\pi \hbar^3 v_F^2} \int_0^{2\pi} d\theta |\hat{U}(q)\sin\theta|^2 \qquad (26)$$

which is identical to Eq. 19. Thus we have demonstrated how to calculate momentum relaxation either directly from BTE, or using the differential scattering cross-section approach.

### IV. Numerical calculation and discussion of mobility and relaxation time in graphene

At this point, we can numerically compute $\tau_m$ with physical parameters for the model in Fig.1. Considering the monolayer *n*-type graphene has an electron density $n = 10^{12} cm^{-2}$, the corresponding Fermi energy $E_F = 0.12 eV$ can be calculated using $n = \frac{1}{2}aE_F^2$ after Eq.15. Assuming zero temperature for convenience in estimation, scattering only occurs at Fermi energy, i.e. $E = E_F$. We use impurity density $n_{imp} = 10^{11} cm^{-2}$, impurity distance $z_0 = 10 nm$, numerical integral can be evaluated that $\int_0^{2\pi} d\theta |\hat{U}(q)\sin\theta|^2 \approx 1.83 \times 10^{-28} eV^2 cm^4$ and thus relaxation time $\tau_m \approx 1.7 \times 10^{-11} s$, or $17 ps$ can be calculated.

With $\tau_m$ known, we can continue to calculate other important electrical parameters such as the carrier mobility and conductivity/resistivity. It is worth mentioning that while conventional semiconductor mobility $\mu$ can be estimated from the Drude model $\mu = \frac{e\tau_m}{m^*}$, where $m^*$ is the effective mass, this expression needs to be modified for massless Dirac electron in graphene. Analogous to the famous $E = mc^2$ relation in relativistic mechanic, we set $m^* = \frac{E_F}{v_F^2}$ and mobility in graphene can be written as

$$\mu = \frac{e v_F^2 \tau_m}{E_F} \tag{27}$$

Plugging $\tau_m = 17 ps$ into Eq. 27, a mobility $\mu \approx 1.43 \times 10^6 cm^2 V^{-1} s^{-1}$ can be calculated, which reveals that graphene has very high carrier mobility, taking account of ionized impurity scattering.

More generally, the energy $E$ in Eq. 26 does not have to be the Fermi energy $E_F$. Using same parameters mentioned above, the potential energy $\hat{U}$ in Fourier space as a function of wave vector $k$ and scattering angle $\theta$ are plotted in Fig. 3a and Fig. 3b.

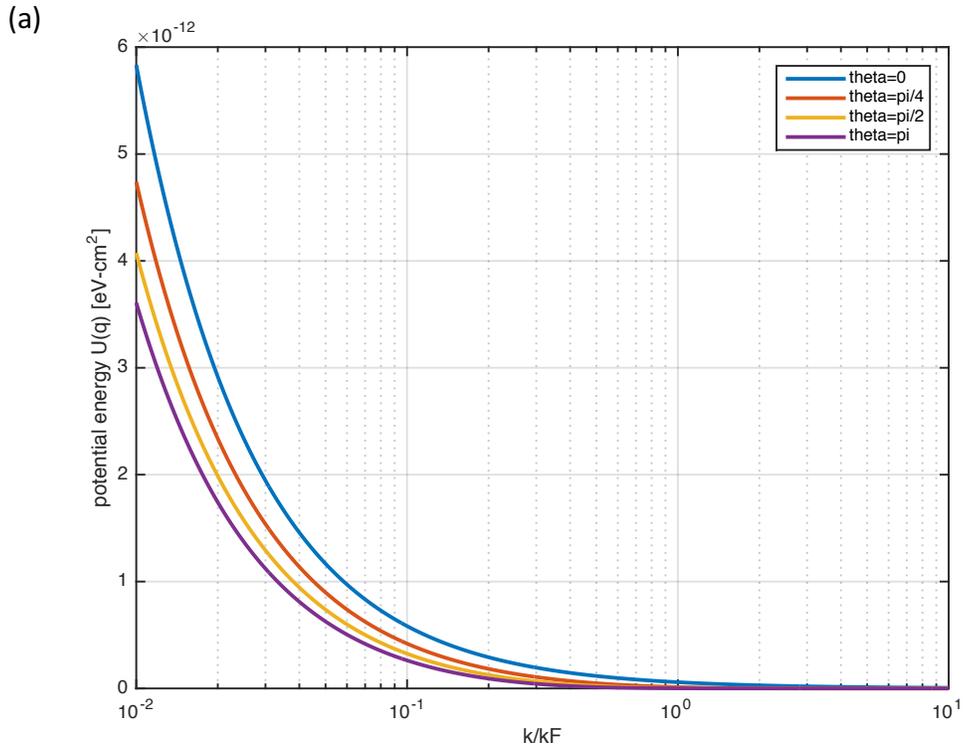

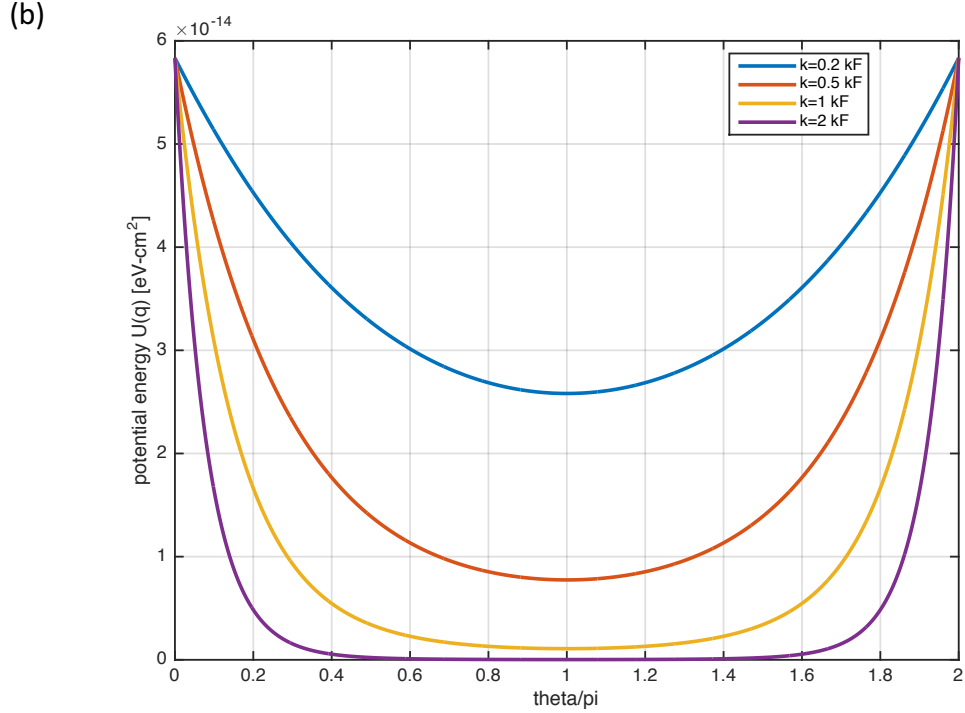

Fig. 3. Potential energy $\hat{U}(q) = \frac{2\pi e^2}{\kappa} \frac{e^{-qz_0}}{q+q_s}$, $q = 2k\sin\frac{\theta}{2}$, $z_0 = 10nm$. (a) $\hat{U}$ as a function of $k$ (normalized to $k_F = \frac{E_F}{\hbar v_F}$) for different scattering angle $\theta$. (b) $\hat{U}$ as a function of $\theta$ (normalized to $\pi$) for different wave vector $k$.

We find that as $k$ increasing, $\hat{U}$ decreases (at a constant $\theta$). For larger $\theta$, $\hat{U}$ decreases faster in Fig. 3a due to the exponential term in Eq. 15. For $\theta = 0$ (forward scattering), the expression reduce to $\hat{U} = \frac{1}{a\hbar v_F k}$ which inversely varies with $k$, since $q = 0$ in this case. For wave vector $k \to 0$, potential energy $\hat{U} \to \infty$.

Calculating the angle integral in Eq. 19 for every $E$, we can plot $\tau_m$ as a function of $E$ in Fig. 4. $\tau_m = 17ps$ can be once again found for $E = E_F$, $z_0 = 10nm$ in the plot. For $z_0 = 10nm$, as $E$ increases, $q$ increases, $\hat{U}(q)$ decreases, and $\tau_m$ first decreases then increases. For low temperature, most scattering events occurs around Fermi energy, therefore it is reasonable to set $E = E_F$ to evaluate relaxation time and mobility. The scattering follows a distribution as $\left(-\frac{\partial f}{\partial E}\right) = \frac{1}{k_B T} f(E) \cdot [1 - f(E)]$, which reduces to delta function for $T = 0K$. For non-zero temperature calculations, please refer to [15]. Additional phonon process can be found in [16-17]. For nonlinear Thomas Fermi screening, please refer to [10].

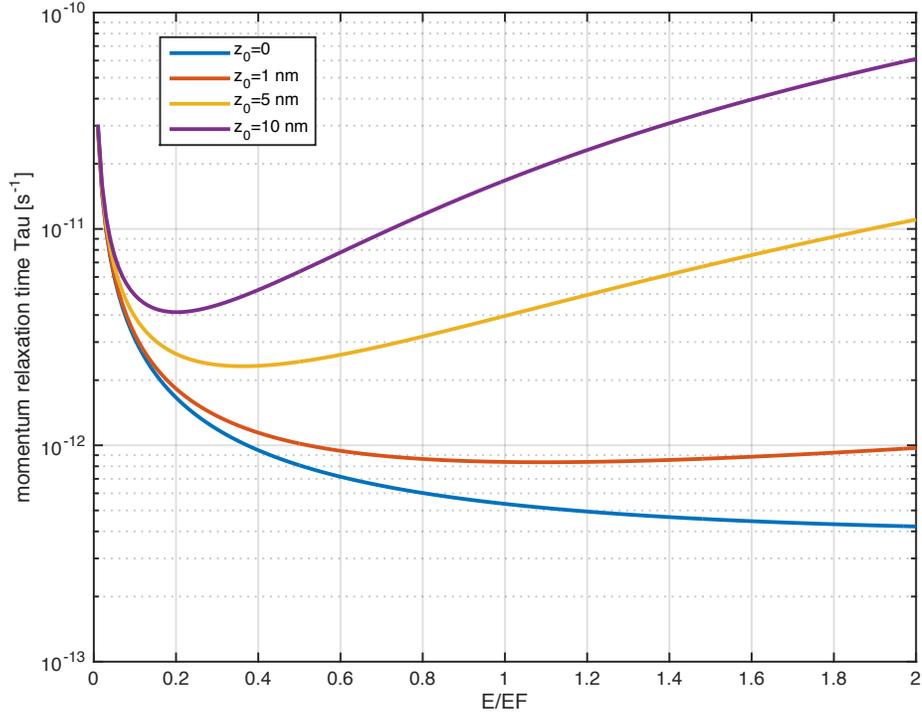

Fig. 4. Momentum relaxation time $\tau_m$ as a function of carrier energy $E$ (normalized to $E_F$) for different impurity distance $z_0$.

In this paper, diffusive transport in graphene with linear Thomas Fermi screening is studied. Ionized impurity scattering is used to illustrate the calculation procedures. Two different methods, directly from Boltzmann transport equation or using scattering cross-section method, are shown to derive the momentum relaxation time. With calculated $\tau_m$, mobility can be further evaluated. As graphene is an isotropic material, $\tau_m$ is solely determined by $E$ (i.e. $\tau_m$ depends on magnitude of incoming $|\vec{k}|$, rather than the angle). Numerically calculated $\hat{U}(q)$ and $\tau_m$ are plotted and discussed in the end.